\documentstyle[aps,epsfig,twocolumn]{revtex}
\begin{document}
\title{Orbital angular momentum of the down converted photons}
\author{Xi-Feng Ren\thanks{%
Electronic address: mldsb@mail.ustc.edu.cn}, Guo-Ping Guo\thanks{%
Electronic address: harryguo@mail.ustc.edu.cn}, Bo Yu, Jian Li, and
Guang-Can Guo}
\address{Key Laboratory of Quantum Information, University of Science and Technology\\
of China, CAS, Hefei 230026, People's Republic of China\bigskip \bigskip }
\maketitle

\begin{abstract}
We calculate the relative amplitude of orbital angular momentum (OAM)
entangled photon pairs from the spontaneous parametric down conversion. The
results show that the amplitude depends on both the two Laguerre indices $l,$
$p$. We also discuss the influences of the mostly used holograms and
mono-mode fibers for mode analyzation. We conclude that only a few
dimensions can be explored from the infinite OAM modes of the down-converted
photon pairs.

PACS number(s): 03.67.Mn, 03.65.Ud, 42.50.Dv
\end{abstract}

\section{Introduction}

Quantum entanglement is a very important property of quantum mechanics. It
is the foundation of quantum teleportation \cite{Benn}, quantum computation 
\cite{shor84,grov97,eke96}, quantum cryptography \cite{eke91}, superdense
coding \cite{ben92}, etc. Up to now most of the theoretical discussions and
experiments are focused on quantum states belonging to two-dimensional
states, or qubits\cite{Sackett00,Pan00,Pan01,Howell02}. In recent years, the
interest in multi-dimensional states, or qudits, is steadily growing for its
promise to realize new types of quantum communication protocols\cite
{Bartlett00,Bechmann00,Bourennane01}, and its properties in quantum
cryptography better than qubits\cite
{Bechmann00,Bourennane01,BechmannA00,Guo02}. These theoretical disscussions%
\cite{Arnaut00,Franke02,Molina02,Barbosa02,Torres03} and experiments\cite
{Arlt99,Mair01,Leach02,Vaziri02,Vaziri03} about multi-dimensional states are
mostly based on the orbital angular momentum (OAM) of the photons. It has
been shown that paraxial Laguerre-Gaussian(LG) laser beams carry a
well-defined orbital angular momentum\cite{Allen92}, and that the LG modes
form a complete Hilbert set. The down-converted photons from spontaneous
parametric down-conversion (SPDC) are entangled in not only polarization, or
spin angular momentum, but also OAM\cite{Mair01,Vaziri02}. This provides a
promise to explore multi-dimensional quantum state in one photon \cite
{Mair01,Molina02}.

In this paper, we calculate the relative amplitude of OAM of the
down-converted photons from SPDC and analyze the possible joint detection
probability under the influence of the experiment elements. Our results show
that the relative amplitude decreases almost exponentially with the growing
of OAM. And only a few dimensions can be explored from the infinite OAM
modes of the down-converted photon pairs. In Section $2$, we briefly
introduce the LG mode, and calculate the relative amplitude of every LG mode
of the down converted photons from SPDC in detail. In theory, the relative
amplitude determines the joint detection probability. However, in practical
experiments, the computer generated holograms and the mono-mode fibres will
inevitably influence the joint detection probability. We discuss the mode
analysis after the computer generated hologram in Section $3$. Section $4$
analyzes the detection efficiency of mono-mode fibre for every LG\ mode.
Section $5$ presents the possible joint detection probability, when the
effects of the holograms and mono-mode fibres are both included. The last
section is the conclusion.

\section{Spontaneous parametric down conversion and OAM}

It is well known that photons can carry both spin angular momentum and OAM%
\cite{Allen92}. Spin angular momentum is associated with polarization and
OAM with the azimuthal phase of the electric field. The normalized LG mode
is given in cylindrical coordinates by

\begin{eqnarray}
LG_p^l(\rho ,\varphi ,z) &=&\sqrt{\frac{2p!}{\pi (\left| l\right| +p)!}}%
\frac 1\omega (\frac{\sqrt{2}\rho }\omega )^{\left| l\right| }L_p^{\left|
l\right| }(\frac{2\rho ^2}{\omega ^2})  \nonumber \\
&&\times \exp (-\rho ^2/\omega ^2)\exp (-ik\rho ^2/2R)  \nonumber \\
&&\exp (-i[2p+\left| l\right| +1]\psi )\exp (-il\varphi ),  \eqnum{1}
\end{eqnarray}
where $L_p^l(x)$ are the associated Laguerre polynomials,

\begin{equation}
L_p^{|l|}(x)=\sum_{m=0}^p(-1)^m\frac{(|l|+p)!}{(p-m)!(|l|+m)!m!}x^m, 
\eqnum{2}
\end{equation}
and the standard definitions for Gaussian beam parameters are used:

$\omega (z)=\omega _0\sqrt{1+(z/z_R)^2}:$ spot size,

$R(z)=z(1+(z_R/z)^2):$ radius of wavefront curvature,

$\psi (z)=\arctan (z/z_R):$ Gouy phase,

$z_R=\frac 12k\omega _0^2:$ Rayleigh range.

$\omega _0$ is the beam width at the beam waist, the index $l$ is referred
to as the winding number, and $(p+1)$ is the number of radial nodes. If the
mode function is a pure LG mode with winding number $l$ , then every photon
of this beam carries an OAM of $l\hbar $. This corresponds to an eigenstate
of the OAM operator with eigenvalue $l\hbar $\cite{Allen92}. If the mode
function is not a pure LG mode, the state is a superposition state, with the
weights dictated by the contribution of the $l$th angular harmonics.

At the beam waist $(z=0)$, the LG mode can be written as

\begin{eqnarray}
LG_p^l(\rho ,\varphi ) &=&\sqrt{\frac{2p!}{\pi (\left| l\right| +p)!}}\frac 1%
{\omega _0}(\frac{\sqrt{2}\rho }{\omega _0})^{\left| l\right| }L_p^{\left|
l\right| }(\frac{2\rho ^2}{\omega _0^2})  \nonumber \\
&&\times \exp (-\rho ^2/\omega _0^2)\exp (-il\varphi ),  \eqnum{3}
\end{eqnarray}

In the following calculation, we use this equation because in the experiment
we always manipulate the light at its beam waist.

In the SPDC process, a thin quadratic nonlinear crystal is illuminated by a
laser pump beam propagating in the $z$ direction, with wave number $k_p$ and
waist $\omega _0$. The generated two-photon quantum state is given by\cite
{Torres03}

\begin{equation}
\left| \Psi \right\rangle
=\sum_{l_1,p_1}\sum_{l_2,p_2}C_{p_1,p_2}^{l_1,l_2}\left|
l_1,p_1;l_2,p_2\right\rangle ,  \eqnum{4}
\end{equation}
where $(l_1,p_1)$ corresponds to the mode of the signal beam and $(l_2,p_2)$
the mode of the idler beam. The probability amplitude $C_{p_1,p_2}^{l_1,l_2}$
is given as\cite{Arnaut00,Franke02,Barbosa02,Torres03}

\begin{equation}
C_{p_1,p_2}^{l_1,l_2}\sim \int dr_{\bot }\Phi (r_{\bot
})[LG_{p_1}^{l_1}(r_{\bot })]^{*}[LG_{p_2}^{l_2}(r_{\bot })]^{*},  \eqnum{5}
\end{equation}
where $r_{\bot }$ is the radial coordinate in the transverse $X-Y$ plane, $%
\Phi (r_{\bot })$ is the spiral distribution of the pump beam at the input
faced of the crystal, $LG_p^l(r_{\bot })$ is the spiral distribution of the
LG mode beam at the same plane.

The weights of the quantum superposition are given by $A_{p_1,p_2}^{l_1,l_2}%
\sim \left| C_{p_1,p_2}^{l_1,l_2}\right| ^2$. It is the ideal joint
detection probability for finding one photon in the signal mode $(l_1,p_1)$
and one photon in the idler mode $(l_2,p_2)$.

Consider the case that the pump beam is in a pure LG mode $LG_{p_0}^{l_0}$
with $p_0=0$. The $LG_0^{l_0}$ mode light at $z=0$ can be written as

\begin{equation}
LG_0^{l_0}(\rho ,\varphi )=\sqrt{\frac 2\pi }\frac 1{\omega _0}(\frac{\sqrt{2%
}\rho }{\omega _0})^{\left| l_0\right| }\exp (-\rho ^2/\omega _0^2)\exp
(-il\varphi ).  \eqnum{6}
\end{equation}
Substitute $LG_0^{l_0}(\rho ,\varphi )$ for $\Phi (r_{\bot })$ into Eq. $(5)$%
, and use the OAM conservation law in SPDC\cite{Arnaut00,Mair01}:

\begin{equation}
l_1+l_2=l_0,  \eqnum{7}
\end{equation}
where $l_1$,$l_2$,$l_0$ are the winding numbers of signal beam, idler beam
and pump beam respectively, we can achieve the probability amplitude

\begin{eqnarray}
C_{p_1,p_2}^{l_1,l_2} &\sim &\sum_{m=0}^{p_1}\sum_{n=0}^{p_2}(\frac 23)^{%
\frac{2m+2n+\left| l_1\right| +\left| l_2\right| +\left| l_0\right| }2%
}(-1)^{m+n}  \nonumber \\
&&\frac{\sqrt{p_1!p_2!(\left| l_1\right| +p_1)!(\left| l_2\right| +p_2)!}}{%
(p_1-m)!(p_2-n)!(\left| l_1\right| +m)!(\left| l_2\right| +n)!m!n!} 
\nonumber \\
&&(\frac{2m+2n+\left| l_1\right| +\left| l_2\right| +\left| l_0\right| }2)!%
\text{.}  \eqnum{8}
\end{eqnarray}

In the case $l_0=0$, or the input beam is Gaussian mode light, Eq. $(8)$ can
be simplified as$(l>0)$

\begin{eqnarray}
C_{p_1,p_2}^{l,-l} &\sim &\sum_{m=0}^{p_1}\sum_{n=0}^{p_2}(\frac 23%
)^{m+n+l}(-1)^{m+n}  \nonumber \\
&&\frac{\sqrt{p_1!p_2!(l+p_1)!(l+p_2)!}(l+m+n)!}{%
(p_1-m)!(p_2-n)!(l+m)!(l+n)!m!n!}.  \eqnum{9}
\end{eqnarray}
It can be easily proved that $%
C_{p_1,p_2}^{l,-l}=C_{p_1,p_2}^{-l,l}=C_{p_2,p_1}^{l,-l}=C_{p_2,p_1}^{-l,l}$.

Table 1 gives the relative value for $p_1,$ $p_2=0,1,2$ and $l=0,1,2$. We
can also illustrate the dependence of the relative probability amplitude $%
C_{p_1,p_2}^{l,-l}$ on $p_1,p_2,$ and $l$, with Fig. 1($l=0$ and $%
p_1,p_2=0,1,2,3,4$) and Fig. 2($p_1=p_2=0$ and $l=0,1,2,3,4$).

From the above table and figures, we can see that the probability amplitude
decreases very rapidly with the growing of $p_1$, $p_2$ and $l$. We then
just consider the cases with $p_1,$ $p_2=0,1,2$ and $l=0,1,2$ when $p_0=$ $%
l_0=0$. In papers\cite{Mair01,Vaziri02}, they also just consider the cases
with $l=0,1,2$.

If additionally assume $p_1=p_2=0$ as in the previous works\cite
{Mair01,Vaziri02,Vaziri03,Torres03}, we can obtain:

\begin{equation}
C_{0,0}^{l,-l}\sim (\frac 23)^l.  \eqnum{10}
\end{equation}
This result can also be achieved from the Eq. $(14)$ of the paper\cite
{Torres03}, when the condition $l_1=-l_2=l$ is assumed.

But till now, this assumption has not been proven either in theoretical
discussions\cite{Arnaut00,Torres03} or in experiments\cite{Mair01,Vaziri02}.
We will discuss the two cases separately in Section $5$. Before proceed, we
analyze the two main experiment elements, computer generated holograms and
mono-mode fibre, which will unavoidably affect the detected relative
probability amplitude.

\section{Computer generated holograms and the mode analysis}

In most of the recent experiments\cite
{Arlt99,Mair01,Leach02,Vaziri02,Vaziri03}, the authors always use computer
generated holograms to transform Gaussian mode light into other LG modes
light, or change the winding number of LG mode light. It is a kind of
transmission holograms with the transmittance function:

\begin{equation}
T(\rho ,\varphi )=\exp (i\delta \frac 1{2\pi }%
%TCIMACRO{\func{mod} }
%BeginExpansion
\mathop{\rm mod}%
%EndExpansion
(l\varphi -\frac{2\pi }\Lambda \rho \cos \varphi ,2\pi )),  \eqnum{11}
\end{equation}
where $\delta $ is the amplitude of the phase modulation, $\Lambda =\frac{%
2\pi }{k_x}$ is the period of the grating at a large distance away from the
fork, $k_x$ is the $x$ component of the simplest reference beam's wave
vector. Corresponding to the diffraction order $m$, the hologram can change
the winding number of the input beam by $\Delta l_m=ml$. The diffraction
efficiency depends on the phase modulation $\delta $. When $\delta =2\pi $,
almost $100\%$ of the incident intensity is diffracted into the first-order.

However, even the input beam is a pure LG mode light, the output beam after
the hologram is not a pure LG mode light. The output light will be the
superposition of the various LG modes with the same $l$, and different $p$%
\cite{Heckenberg92}. In addition, the beam waists of the input and output
beam will affect the weights of different components of the output beam.
Assume the input beam and the output beam have the same waists $\omega _0$,
thus the complex expansion coefficients of the decomposition of the $m$th
diffraction order can be calculated as:

\begin{eqnarray}
a_{pl} &=&\int \int (LG_{p_1^{\prime }}^{l_1^{\prime }}(\rho ,\varphi )\exp
(-im\frac{2\pi }\Lambda r\cos \varphi ))^{*}  \nonumber \\
&&T(\rho ,\varphi )E_{in}(\rho ,\varphi )\rho d\rho d\varphi ,  \eqnum{12}
\end{eqnarray}
where $E_{in}(\rho ,\varphi )=LG_{p_1}^{l_1}(\rho ,\varphi )$. Consider the
first-order diffraction, or $m=1$, Eq. $(12)$ can be then rewritten as

\begin{equation}
a_{p_1,p_1^{\prime }}^{l_1,l_1^{\prime }}=\int \int (LG_{p_1^{\prime
}}^{l_1^{\prime }}(\rho ,\varphi ))^{*}\exp (-i\Delta l\varphi
))LG_{p_1}^{l_1}(\rho ,\varphi )\rho d\rho d\varphi ,  \eqnum{13}
\end{equation}
where $\Delta l=l_1-l_1^{\prime }$ is the winding number changed by the
hologram. The relative weight of the output modes is given by

\begin{equation}
P_{p_1,p_1^{\prime }}^{l_1,l_1^{\prime }}=\left| a_{p_1,p_1^{\prime
}}^{l_1,l_1^{\prime }}\right| ^2.  \eqnum{14}
\end{equation}

As the mono-mode fibres can only detect the photons with $l=0$, we consider
the case that the output light is in the modes with $l_1^{\prime }=0$. Then $%
P_{p_1,p_1^{\prime }}^{l_1,l_1^{\prime }}$ can be simplified as $%
P_{p_1,p_1^{\prime }}^{\Delta l}$, and $P_{p_1,p_1^{\prime }}^{\Delta
l}=P_{p_1,p_1^{\prime }}^{-\Delta l}$. In most of the experiments\cite
{Mair01,Leach02,Vaziri02,Vaziri03}, only the holograms of $\Delta l=1$ or $2$
are employed. Table 2 and Table 3 give the relative weight of different
modes after the computer generated hologram with $\Delta l=1$ and $2$.

From Table 2 and Table 3, we can see most of the input mode $LG_p^{\Delta l}$
is converted into $LG_p^0$ and $LG_{p+1}^0$. Thereby, we only consider the
case that the output light is in the modes with $p^{\prime }=0,1,2,3$.

\section{Mono-mode fibre and detection efficiency of the OAM modes}

It is known that only one mode of light can transmit in the mono-mode fibre: 
$HE_{11}$ mode. And in practical calculation, we always use Gaussian mode to
replace the $HE_{11}$ mode. The Gaussian mode is

\begin{equation}
E(\rho )=E(0)\exp (-\rho ^2/\omega ^2),  \eqnum{15}
\end{equation}
where $d=2\omega $ is the Mode Field Diameter(MFD) of the fibre, $E(0)$ is
amplitude of field at the fibre center. For $LG_p^l$ mode light, the
detection efficiency is given as

\begin{equation}
Q_{l,p}=\frac{(\int \int (LG_p^l)^{*}E(\rho )\rho d\rho d\varphi )^2}{\int
\int (LG_p^l)^{*}LG_p^l\rho d\rho d\varphi \int \int E(\rho )^{*}E(\rho
)\rho d\rho d\varphi }.  \eqnum{16}
\end{equation}

Obviously if $l\neq 0$, then $Q_{l,p}=0$. For the case $l=0$, Eq. $(16)$ can
be simplified as:

\begin{equation}
Q_p=\frac{(\int \int (LG_p^0)^{*}E(\rho )\rho d\rho d\varphi )^2}{\int \int
(LG_p^0)^{*}LG_p^0\rho d\rho d\varphi \int \int E(\rho )^{*}E(\rho )\rho
d\rho d\varphi }.  \eqnum{17}
\end{equation}

To calculate the relative joint detection probability of the down-converted
photons from SPDC, we only need the relative detection efficiency of the $%
LG_0^0$, $LG_1^0$ , $LG_2^0$ and $LG_3^0$, but not their absolute
efficiency. Assume the waist size of the input beam is adjusted equal to $%
d/2 $. Then when the detection area is much more larger than the
cross-section of the input light , only $LG_0^0$ mode light can be detected.
But in practice, the detection area is determined by the fibre diameter. To
simplify calculation, we further assume that the detection area is a round
area with diameter equal to the MFD. Thus the integral for $\rho $ is from $%
0 $ to $\omega $. With these assumptions, the relative efficiencies for $%
p=0,1,2,3$ can be written as

\begin{equation}
Q_0:Q_1:Q_2:Q_3=1:0.263:0:0.036  \eqnum{18}
\end{equation}

\section{Relative joint detection probabilities of OAM entangled photons
from SPDC}

With the above discussions about the influence of computer generated
holograms and mono-mode fibres, we now consider the joint detection
probability for OAM\ entangled photons generated from an experimental set-up
similar to the work\cite{Mair01}. The joint detection probability can be
written as

\begin{equation}
R_l=\sum_{p_1=0}^2\sum_{p_2=0}^2((C_{p_1,p_2}^{l,-l})^2\sum_{p_1^{^{\prime
}}=0}^{p_1+1}\sum_{p_2^{^{\prime }}=0}^{p_2+1}(P_{p_1,p_1^{^{\prime
}}}^{-l}P_{p_2,p_2^{^{\prime }}}^lQ_{p_1^{^{\prime }}}Q_{p_2^{^{\prime }}}))%
\text{.}  \eqnum{19}
\end{equation}
When $l=0$, $R_0$ gives the joint detection probability that there is no
hologram in both the signal and idler beam. And for the case $l\neq 0$, $R_l$
represents the joint detection probability with one $\Delta l=-l$ hologram
in the signal beam and one $\Delta l=l$ hologram in the idler beam.
Obviously, $R_l=R_{-l}$.

Substitute the values of $C$, $P$ and $Q$ calculated in the above sections
to Eq. $(19)$, we can get the relative joint detection probability of the
three cases $l=0,1,2$ as

\begin{equation}
R_0:R_1:R_2=1:0.346:0.101.  \eqnum{20}
\end{equation}

If we also assume that $p_1=p_2=0$ for the SPDC process as the recent papers%
\cite{Mair01,Vaziri02,Vaziri03,Torres03}, the joint detection probability
can thus be written as 
\begin{equation}
R_l=((C_{0,0}^{l,-l})^2\sum_{p_1^{^{\prime }}=0}^1\sum_{p_2^{^{\prime
}}=0}^1(P_{0,p_1^{^{\prime }}}^{-l}P_{0,p_2^{^{\prime }}}^lQ_{p_1^{^{\prime
}}}Q_{p_2^{^{\prime }}}))\text{.}  \eqnum{21}
\end{equation}
And the relation for $l=0,1,2$ becomes:

\begin{equation}
R_0:R_1:R_2=1:0.311:0.079.  \eqnum{22}
\end{equation}
The difference between Eq. $(20)$ and $(22)$ is caused by the additional
assumption that Laguerre index $p_1=p_2=0$. But this assumption has not been
proven either in theoretical discussions\cite{Arnaut00,Torres03} or in
experiments\cite{Mair01,Vaziri02}. From the above calculations, we can see
that this assumption will cause non-trivial influence to the relative joint
detection probability. Our results put forward a feasible method to verify
the assumption.

If we rule out the influence of the holograms and mono-mode fibres, and make
the assumption of $p_1=p_2=0$, the relationship for the relative joint
detection probability of the cases $l=0,1,2$ becomes

\begin{equation}
R_0:R_1:R_2=1:0.444:0.198,  \eqnum{23}
\end{equation}
Compare Eq. $(22)$ with Eq. $(23)$, we can see that the experimental
elements can apparently influence the joint detection probability.

In the experiment by Vaziri and co-workers\cite{Vaziri02}, they found that
the state of the OAM entangled photons from SPDC was given by

\begin{equation}
\psi =0.65\left| 0,0\right\rangle +0.60\left| 1,-1\right\rangle +0.47\left|
-1,1\right\rangle .  \eqnum{24}
\end{equation}
From this equation we can get the relationship for the relative joint
detection probabilities of the cases $l=0,1,-1$ as

\begin{equation}
R_0:R_1:R_{-1}=1:0.852:0.523.  \eqnum{25}
\end{equation}
Including the influence of computer generated holograms and mono-mode
fibres, and loosening the assumption of $p_1=p_2=0$, we expect this relation
be 
\begin{equation}
R_0:R_1:R_{-1}=1:0.346:0.346.  \eqnum{26}
\end{equation}
The reason for the difference between Eq. $(25)$ and $(26)$ might be as
follows: in experiment, the diffraction efficiency of the hologram can not
be $100\%$ . Generally, different holograms have different diffraction
efficiencies. The waists of the input beam and the output beam of holograms
will also affect the final experiment detection probabilities. In addition,
the fibre diameter and MFD of the practical mono-mode fibre will also affect
the detection efficiencies of different modes light. Evidently, the
detection efficiency of avalanche detectors has little effect on the
relative joint detection probabilities. Thus in the practical experiment,
the $P$ and $Q\ $values have to be adjusted according to the particular
conditions.

\section{Conclusion}

In conclusion, we have calculated the probability amplitudes of different LG
modes of the down converted photons. Our results show that the relative
amplitude decreases almost exponentially with growing of OAM. We also
discussed the impact of the previous assumption for $p$ on the joint
detection probability. In addition, we analyzed the influences of the
experiment elements. We concluded that only a few dimensions can be explored
from the infinite OAM modes of the down-converted photon pairs. The
experiment verification of the present theory is straightforward and is
currently in progress in our laboratory.

\begin{center}
{\bf Acknowledgments}
\end{center}

This work was funded by the National Fundamental Research Program
(2001CB309300), the Innovation Funds from Chinese Academy of Sciences, and
also by the outstanding Ph. D thesis award and the CAS's talented scientist
award rewarded to Lu-Ming Duan.

{\bf Table 1. }The relative probability amplitude $C_{p_1,p_2}^{l,-l}$ of
the down converted photons from SPDC. We let $C_{0,0}^{0,0}$ be unity.

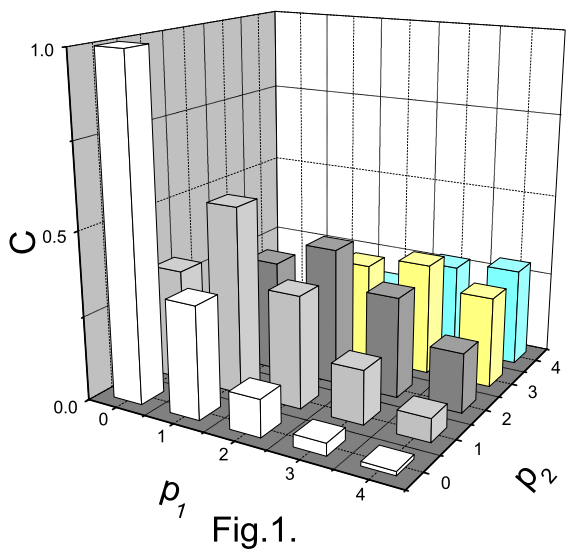
{\bf Fig. 1.} The relative probability amplitude $C$ for $l=0$ and $%
p_1,p_2=0,1,2,3,4$. We let $C_{0,0}^{0,0}$ be unity.

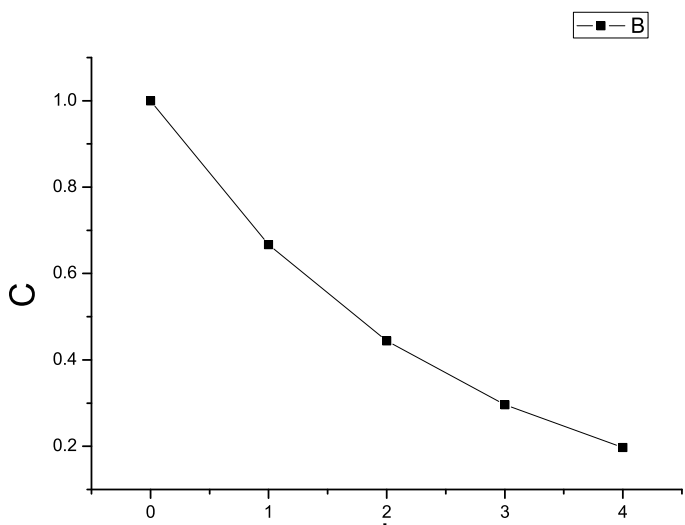
{\bf Fig. 2.} The relative probability amplitude $C$ for $p_1=p_2=0$ and $%
l=0,1,2,3,4$. We let $C_{0,0}^{0,0}$ be unity.

{\bf Table 2. }The relative weight of different modes after the computer
generated hologram with $\Delta l=1$.

{\bf Table 3. }The relative weight of different modes after the computer
generated hologram with $\Delta l=2$.


\begin{references}
\bibitem{Benn}  C. H. Bennett {\it et al}., Phys. Rev. Lett. {\bf 70}, 1895
1993.

\bibitem{shor84}  P. W. Shor, in {\it Proceeding of the 35th Annual
Symposium on the Foundation of Computer Science}, p. 124-133 (IEEE Computer
Science Press, Los Alamitos, California, 1994).

\bibitem{grov97}  L. K. Grover, Phys. Rev. Lett. {\bf 79}, 325 1997.

\bibitem{eke96}  A. K. Ekert and R. Jozsa, Rev. Mod. Phys. {\bf 68}, 733
1996.

\bibitem{eke91}  A. K. Ekert, Phys. Rev. Lett. {\bf 67}, 661 1991.

\bibitem{ben92}  C. H. Bennett and S. J. Wiesner, Phys. Rev. Lett.{\bf \ 69}%
, 2881 1992.

\bibitem{Sackett00}  C.A. Sackett, D. Kielpinski, B.E. King, C.Langer, V.\
Meyer, C.J. Myatt, M. Rowe, Q.A. Turchette, W.B. Itano, D.J.\ Wineland, and
C. Monroe, Nature(London) {\bf 404}, 256-259 2000.

\bibitem{Pan00}  J.-W Pan, D. Bouwmeester, M. Daniell, H. Weinfurter, and A.
Zeilinger, Nature (London) {\bf 403}, 515 2000.

\bibitem{Pan01}  J.-W Pan, M. Daniell, S. Gasparoni, G. Weihs, and A.
Zeilinger, Phys.Rev.Lett. {\bf 86}, 4435-4438 2001.

\bibitem{Howell02}  John C. Howell, Antia Lamas-Linares, and Dik
Bouwmeester, Phys.Rev.Lett. {\bf 88}, 030401 2002.

\bibitem{Bartlett00}  S. D. Bartlett, H. de Guise, and B. C. Sanders.
quant-ph/0011080, 2000.

\bibitem{Bechmann00}  H. Bechmann-Pasquinucci and A. Peres. Phys.Rev.Lett. 
{\bf 85}, 3133, 2000.

\bibitem{Bourennane01}  M. Bourennane, A. Karlsson, and G. Bj\"{o}rk.
Phys.Rev.A. {\bf 64}, 012306, 2001.

\bibitem{BechmannA00}  H. Bechmann-Pasquinucci and W. Tittel. Phys.Rev.A. 
{\bf 61}, 62308, 2000.

\bibitem{Guo02}  Guo-Ping Guo, {\it et. al. }Phys.Rev.A. {\bf 64}, 042301,
2001.

\bibitem{Arnaut00}  H. H. Arnaut and G. A. Barbosa, Phys.Rev.Lett. {\bf 85},
286, 2000.

\bibitem{Franke02}  S. Franke-Arnold, S. M. Barnett, M.\ J. Padgett, and L.
Allen, Phys.Rev.A. {\bf 65}, 033823, 2002.

\bibitem{Molina02}  G. Molina-Terriza, J. P. Torres, and L. Torner,
Phys.Rev.Lett. {\bf 88}, 013601, 2002.

\bibitem{Barbosa02}  G. A. Barbosa and H. H. Arnaut, Phys.Rev.A. {\bf 65},
053801, 2002.

\bibitem{Torres03}  J. P. Torres, Y. Deyanova, and L. Torner, Phys.Rev.A. 
{\bf 67}, 052313, 2003.

\bibitem{Arlt99}  J. A. Arlt, K. Dholakia, L. Allen, and M. J. Padgett,
Phys.Rev.A. {\bf 59}, 3950, 1999.

\bibitem{Mair01}  A. Mair, A. Vaziri, G. Weihs, and A. Zeilinger, Nature
(London) {\bf 412}, 313, 2001.

\bibitem{Leach02}  J. Leach, M. J. Padgett, S. M. Barnett, S. Franke-Arnold,
and J. Courtial, Phys.Rev.Lett. {\bf 88}, 257901, 2002.

\bibitem{Vaziri02}  A. Vaziri, G. Weihs, and A. Zeilinger, Phys.Rev.Lett. 
{\bf 89}, 240401, 2002.

\bibitem{Vaziri03}  A. Vaziri, J.-W Pan, T. Jenewein, G. Weihs, and A.
Zeilinger, quant-ph/0303003, 2003.

\bibitem{Allen92}  L. Allen, M. W. Beijersbergen, R. J. C. Spreeuw, and J.
P. Woerdman, Phys.Rev.A. {\bf 45}, 8185, 1992.

\bibitem{Heckenberg92}  N. R. Heckenberg, R. McDuff, C. P. Smith, H.
Rubinsztein-Dunlop, and M. J. Wegener, Opt. Quant. Elec, {\bf 24}, S951,
1992.
\end{references}
\end{document}